\documentclass{article}
\usepackage{graphicx} 
\usepackage{graphics}

\usepackage{amsmath}   
\usepackage{amsfonts}
\usepackage{amssymb}

\newcommand{\dfr}{\widehat{\rm d}} 

\renewcommand\baselinestretch 2
\begin{document}
%
%
\def\l{\left}
\def\r{\right}
\def\beq{\begin{equation}}
\def\eeq{\end{equation}}
\def\d{\partial}

\title{Emergence of complex and spinor wave functions in Scale Relativity. II. Lorentz invariance and bi-spinors}
\author{Marie-No\"elle C\'el\'erier and Laurent Nottale, \footnote{LUTH, Observatoire de Paris, CNRS, Universit\'e Paris-Diderot,
5 place Jules Janssen, 92195 Meudon Cedex, France;
e-mails: marie-noelle.celerier@obspm.fr, laurent.nottale@obspm.fr }}

\maketitle

\abstract

Owing to the non-differentiable nature of the theory of Scale Relativity, the emergence of complex wave functions, then of spinors and bi-spinors occurs naturally in its framework. The wave function is here a manifestation of the velocity field of geodesics of a continuous and non-differentiable (therefore fractal) space-time. In a first paper (Paper I), we have presented the general argument which leads to this result using an elaborate and more detailed derivation than previously displayed. We have therefore been able to show how the complex wave function emerges naturally from the doubling of the velocity field and to revisit the derivation of the non relativistic Schr\"odinger equation of motion. In the present paper (Paper II) we deal with relativistic motion and detail the natural emergence of the bi-spinors from such first principles of the theory. Moreover, while Lorentz invariance has been up to now inferred from mathematical results obtained in stochastic mechanics, we display here a new and detailed derivation of the way one can obtain a Lorentz invariant expression for the expectation value of the product of two independent fractal fluctuation fields in the sole framework of the theory of Scale Relativity. These new results allow us to enhance the robustness of our derivation of the two main equations of motion of relativistic quantum mechanics (the Klein-Gordon and Dirac equations) which we revisit here at length.




\section{Introduction}
\label{int}


From a physical point of view, the theory of Scale Relativity (SR) is the generalization to scales of the relativity principle of motion which underlies the foundation of a large part of classical physics. From a mathematical point of view, it is the giving up of space-time differentiability. Both generalizations result in the fractal nature of space-time \cite{LN93}.

One of the main achievements of this theory is the foundation of quantum mechanics on first principles. In its framework, the quantum mechanical postulates have been derived and the complex, then spinorial, then bi-spinorial nature of the wave function has been naturally recovered \cite{LN93,NC07,CN06,CN03,CN04}, while the corresponding quantum mechanical motion equations, the Schr\"odinger \cite{LN93}, Pauli \cite{CN06}, Klein-Gordon \cite{LN94,LN96} and Dirac \cite{CN03,CN04} equations have been demonstrated.

The theory has also allowed us to generalize the quantum mechanical motion equations to the macrocopic realm. This is obtained when the ratio of the reduced Planck constant $\hbar$ over twice the electron mass, operating in the microscopic domain, is given a more general interpretation in terms of a macroscopic constant ${\cal D}$ whose value is linked to the physical system under study. We have been therefore able to derive a macroscopic Schr\"odinger-like equation with numerous applications in physics and other fields (see, e.g., Ref. \cite{LN11} for a review) and a macroscopic Dirac-like equation whose non-relativist limit might tentatively reproduce some turbulent fluid behavior \cite{MNC09}.

The emergence of the complex numbers and their generalizations, the quaternions, and the bi-quaternions, is issued from the successive doublings of the velocity fields due to the non-differentiability of the fractal functions representing the space-(time) coordinates. These doublings have been extensively studied in previous works (see, e.g., Ref. \cite{LN11} for a review).

Now, a new and more detailed derivation of the way the two-valuedness of the mean velocity field emerges naturally in terms of a (+) and (-) velocity has been put forward in Ref. \cite{NC13} (Paper I). It yields $V$ and $U$ velocity fields, which are subsequently combined in terms of the complex velocity field ${\cal V} = V- i U$ used in the geodesic equation $\widehat{\rm d}{\cal V}/{\rm d}t=0$ written in the fractal three-dimensional space and from which the Schr\"odinger equation is derived. We are led here to generalize these results to the four-dimensional space-time, being therefore allowed to recover the complex 4-velocity field needed to obtain the Klein-Gordon equation.

We extend then this two-valuedness to successive other doublings. We obtain thus naturally a more symmetric expression than that provided in Refs. \cite{CN03} and \cite{CN04} for the bi-quaternionic velocity field yielding the bi-spinor and the Dirac equation, from which the spinor and the Pauli equation have been derived \cite{CN06}.

Another essential tool used to derive the quantum mechanical motion equations is the  expectation value of the product of two independent fractal fluctuation fields. Its expression appears most naturally in the non-relativistic three-dimensional generalization of the theory, then in its four-dimensional relativistic version. Due to the stochastic nature of these fluctuations, the transition from non-relativistic to relativistic motion allows us to write a Lorentz invariant expression for this expectation value in the relativistic case and therefore to obtain the Klein-Gordon and Dirac equations. The transition from non-relativistic motion to the relativistic Klein-Gordon equation has been first implemented in the framework of stochastic mechanics \cite{GR78,MS88,TZ90} and the mathematical tools thus provided have been applied to the scale relativistic fractal fluctuations allowing us to obtain the relativistic equations of quantum mechanics \cite{CN03,CN04,LN94,LN96}. We propose here a new derivation of the way one can obtain this Lorentz invariance in the sole framework of SR.

The structure of the present paper is as follows. In Sec.~\ref{Li} we generalize the results of Paper I \cite{NC13} to the full space-time of the relativistic case and obtain naturally the emergence of Lorentz invariance in the framework of SR, allowing us to give in Sec.~\ref{vf} a more accurate derivation of the relativistic Klein-Gordon and Dirac equations. Section \ref{concl} is devoted to the conclusions.


\section{Lorentz invariance of the relativistic tools of SR}
\label{Li}


We begin by recalling below the main equations displayed in Paper I \cite{NC13} which we will need here for our purpose. It has been shown in Paper I \cite{NC13} that the differential element ${\rm d} X(x(t),t,{\rm d} t,\varepsilon_t)$ used in the definition of a fractal velocity field $V(x(t),t,{\rm d} t,\varepsilon_t) ={\rm d} X/{\rm d} t$ considered on a fractal curve, with the time coordinate $t$ as a parameter along this curve, can be decomposed into a differentiable linear contribution and a non-linear fractal fluctuation, such as
\beq
{\rm d} X= V[x(t),t] {\rm d} t +  U[x(t),t] \varepsilon_t + {\rm d} \xi.
\eeq
The fractal fluctuation, ${\rm d} \xi$, has been there described by a stochastic variable which possesses the nature of a space resolution  $\varepsilon_x>0$, i.e.,
\beq
{\rm d} \xi= \eta \, \varepsilon_x,
\eeq
where $\eta$ is a dimensionless variable, normalized according to $\langle \eta \rangle=0$ and $\langle \eta^2 \rangle =1$.
When the fractal dimension exhibits the critical value $D_f=2$ \cite{LN93,LN96,FH65}, the space and time resolutions are related by
\beq
\varepsilon_x = \sqrt{2 {\cal D} \, \varepsilon_t}.
\eeq
We have therefore obtained
\beq
{\rm d} X= V[x(t),t] {\rm d} t +  U[x(t),t] \varepsilon_t +  \eta \sqrt{2 {\cal D} \, \varepsilon_t}.
\eeq
On the interface $\varepsilon_t = |{\rm d} t|$, which defines the physical fractal derivative, this becomes
\beq
{\rm d} X= V {\rm d} t +  U |{\rm d} t| +  \eta \sqrt{2 {\cal D} \, |{\rm d} t|}.
\eeq
Two possibilities occur therefore for the elementary displacements
\beq
{\rm d} t>0: \quad |{\rm d} t|={\rm d} t, \quad {\rm d} X_+ = (V +  U) {\rm d} t +  \eta \sqrt{2 {\cal D} \, {\rm d} t}.
\eeq
\beq
{\rm d} t<0: \quad |{\rm d} t|=-{\rm d} t, \quad {\rm d} X_- = (V - U) {\rm d} t +  \eta \sqrt{-2 {\cal D} \, {\rm d} t}.
\eeq
By setting $v_+=V+U$ and $v_-=V-U$, the two-valuedness of the mean velocity field has been recovered in the non-relativistic case described in Paper I \cite{NC13}.

We consider now the relativistic case and derive a relativistically covariant expression for the expectation value of the product of two independent fractal fluctuation fields  needed to complete the derivation of the Klein-Gordon and Dirac equations in SR.

The generalization of the results displayed in Paper I \cite{NC13} to a four-dimensional space-time runs as follows. The curvilinear parameter along the geodesics is now the invariant $s$. Not only space, but the whole continuous space-time is assumed to be non-differentiable, hence fractal. We can write the elementary displacement, ${\rm d} X^{\mu}$, for each coordinate $x^\mu$, $\mu = 0,1,2,3$, along a geodesic of fractal dimension $D_f = 2$ as
\begin{equation}
{\rm d} X^{\mu} = V^{\mu}[x^{\mu}(s),s] {\rm d} s + U^{\mu}[x^{\mu}(s),s] \epsilon_s + {\rm d} \xi^{\mu},
\label{dX4D}
\end{equation}
with
\begin{equation}
{\rm d} \xi^{\mu} = \eta_{\mu} \epsilon_{x^{\mu}},
\label{dxi4D1}
\end{equation}
where the $\eta^{\mu}$'s are dimensionless variables normalized according to $\langle \eta^{\mu} \rangle = 0$, $\langle \eta^{\mu} \eta^{\nu} \rangle = \delta^{\mu \nu}$, and
\begin{equation}
\epsilon_{x^{\mu}} = \sqrt{2 {\cal D} \, \epsilon_s}.
\label{eps4D}
\end{equation}
At the interface where the derivatives with respect to $s$ take their physical values whatever the scale, we make the particularly meaningful choice $\epsilon_s = |{\rm d}s|$ which implies
\begin{equation}
{\rm d} \xi^{\mu} = \eta^{\mu} \sqrt{2 {\cal D} \, \epsilon_s} =  \eta^{\mu}  \sqrt{2 {\cal D} \, |{\rm d}s|}.
\label{dxi4D2}
\end{equation}
In the case ${\rm d} s > 0$ {\it for each $\mu$}, we thus obtain
\begin{equation}
{\rm d} \xi^{\mu}_+ = \eta^{\mu}  \sqrt{2 {\cal D} \, {\rm d}s},
\label{dxi4D+}
\end{equation}
and in the case ${\rm d} s < 0$ {\it for each $\mu$},
\begin{equation}
{\rm d} \xi^{\mu}_- = \eta^{\mu}  \sqrt{- 2 {\cal D} \, {\rm d}s}.
\label{dxi4D-}
\end{equation}
This implies
\begin{equation}
\langle {\rm d} \xi^{\mu}_{\pm} {\rm d} \xi^{\nu}_{\pm} \rangle = \pm 2 {\cal D} \delta ^{\mu \nu} {\rm d} s.
\label{dxipm2}
\end{equation}

We consider now that each ${\rm d} \xi^{\mu}$ behaves such as an independent fractal fluctuation around its own mean equal to zero, in accordance with $\langle \eta^{\mu} \rangle = 0$. This justifies to consider it as a stochastic process which we choose to be of Berstein type, since such a type includes a large class of stochastic processes among which Markov and anti-Markov processes. We inspire ourselves here from the reasonings put forward by Guerra and Ruggiero \cite{GR78} and Serva \cite{MS88} in the framework of a stochastic diffusion theory, while adapting them to the scale relativistic concepts.

The reasoning proposed in Ref. \cite{GR78} takes into account the fact that time and space play formally an identical role in relativistic theories. The authors of Ref. \cite{GR78} consider therefore that a particle can change direction not only in space, but also in time. A particle moving backward in time can be interpreted as an antiparticle and the point of inversion of its temporal motion can be interpreted as a point where pairs of particles are created or annihilated. Generally, the trajectory of a particle can intersect many times a hyper-surface of constant time coordinate $t$. The theory can thus be thought of as describing a many particle system with creation and destruction of pairs. A possible interpretation is that the four-dimensional one-particle process is merely a mathematical ``trick'' used to describe a many particle system. The invariant $s$ has therefore no physical meaning. It is merely a tool used to associate a probability measure to any possible realization of the many particle process.

Another possible interpretation, put forward in Ref. \cite{MS88} and mathematically equivalent to the previous one, is that the one-particle trajectories possess a physical meaning and that a microscopic particle can actually invert its temporal motion. The invariant parameter $s$ is here still a tool used to associate a probability to each realization but it also appears as a generalization of the proper time.

We want to stress now that, at variance with the method developed in Refs. \cite{GR78} and \cite{MS88}, we do not consider here the inversion of the invariant parameter $s$, but the breaking of the symmetry ${\rm d} s \leftrightarrow -\,{\rm d} s$. However, as regards the above interpretation in terms of backward motion on the fractal geodesics considered here in a local sense and creation and annihilation of pairs, an inversion of ${\rm d} s$ is physically equivalent to an inversion of the parameter $s$ itself. This has already been stressed in Ref. \cite{LN11}. Following a reasoning inspired by Serva \cite{MS88}, we consider two classes of stochastic processes that both contain a combination of Markovian and anti-Markovian processes and can be thought of as Berstein processes: the first with a positive increment of the $s$ parameter for the time coordinate and negative increments of this $s$ parameter for the space coordinates, the second with a negative increment of the $s$ parameter for the time coordinate and positive increments of this $s$ parameter for the space coordinates.

We choose thus, for the first process denoted 1, ${\rm d} s > 0$ for the $X^0_1$ coordinate, which gives
\begin{equation}
{\rm d} X^0_1 = V^0_1 {\rm d} s + U^0_1 {\rm d} s + \eta^0_1  \sqrt{2 {\cal D} {\rm d} s},
\label{dX01}
\end{equation}
which we write
\begin{equation}
{\rm d} X^0_- = v^0_- {\rm d} s + \eta^0_-  \sqrt{2 {\cal D} {\rm d}s} = v^0_- {\rm d} s + {\rm d} \xi^0_-,
\label{dX0m}
\end{equation}
with $v^0_- = V^0_1 + U^0_1$. And we choose conversely, for this process 1, ${\rm d}s < 0$ for the $X^i_1$ coordinates, which yields
\begin{equation}
{\rm d} X^i_1 = V^i_1 {\rm d} s - U^i_1 {\rm d} s + \eta^i_1  \sqrt{- 2 {\cal D} {\rm d} s},
\label{dXi1}
\end{equation}
which we write
\begin{equation}
{\rm d} X^i_- = v^i_- {\rm d} s + \eta^i_-  \sqrt{- 2 {\cal D} {\rm d}s} = v^i_- {\rm d} s + {\rm d} \xi^i_-,
\label{dXim}
\end{equation}
with $v^i_- = V^i_1 - U^i_1$. The upper index 0 denotes here the time coordinate, while $i = 1,2,3$ denotes the spatial coordinates.

For the second process denoted 2, we choose ${\rm d}s < 0$ for the $X^0_2$ coordinate, which gives
\begin{equation}
{\rm d} X^0_2 = V^0_2 {\rm d} s - U^0_2 {\rm d} s + \eta^0_2  \sqrt{- 2 {\cal D} {\rm d} s},
\label{dX02}
\end{equation}
which we write
\begin{equation}
{\rm d} X^0_+ = v^0_+ {\rm d} s + \eta^0_+  \sqrt{- 2 {\cal D} {\rm d}s} = v^0_+ {\rm d} s + {\rm d} \xi^0_+,
\label{dX0p}
\end{equation}
with $v^0_+ = V^0_2 - U^0_2$. And we choose conversely, for this process 2, ${\rm d}s > 0$ for the $X^i_2$ coordinates, which yields
\begin{equation}
{\rm d} X^i_2 = V^i_2 {\rm d} s + U^i_2 {\rm d} s + \eta^i_2  \sqrt{ 2 {\cal D} {\rm d} s},
\label{dXi2}
\end{equation}
which we write
\begin{equation}
{\rm d} X^i_+ = v^i_+ {\rm d} s + \eta^i_+  \sqrt{2 {\cal D} {\rm d}s} = v^i_+ {\rm d} s + {\rm d} \xi^i_+,
\label{dXip}
\end{equation}
with $v^i_+ = V^i_2 + U^i_2$.

Since $\langle \left( \eta^0_+ \right)^2 \rangle = \langle \left( \eta^0_- \right)^2 \rangle = \langle \left( \eta^i_+ \right)^2 \rangle = \langle \left( \eta^i_- \right)^2 \rangle = 1$ from our normalization choice, we obtain from Eqs.~(\ref{dX0m}), (\ref{dXim}), (\ref{dX0p}) and (\ref{dXip})
\begin{equation}
\langle {\rm d} \xi^{\mu}_{\pm} {\rm d} \xi^{\nu}_{\pm} \rangle = \mp 2 \eta^{\mu \nu} 
{\cal D} {\rm d} s,
\label{dxidxi}
\end{equation}
where $\eta^{\mu \nu}$ is the Minkowski metric with signature $(+ - - -)$. We have therefore demonstrated, {\it in the framework of SR}, how one can pass from Eq.~(\ref{dxipm2}) to Eq.~(\ref{dxidxi}) and obtain a relation between the fractal fluctuations which implements the Lorentz invariance of the theory. Note that Eq.~(\ref{dxidxi}) has been, up to know, used at length for the derivation of the relativistic motion equations of quantum mechanics, obtained with $\hbar = 2 m c {\cal D}$ \cite{CN03,CN04,LN94,LN96}, and that a thorough derivation of this equation in the context of SR should be indeed welcome.


\section{The velocity fields of the relativistic quantum mechanical equations of motion}
\label{vf}


\subsection{The velocity field of the Klein-Gordon equation}
\label{KG}

To recover the complex nature of the velocity field ${\cal V}^{\mu}$ used for the derivation of the Klein-Gordon equation, we use the following relabeling already displayed in Eqs.~(\ref{dX0m}), (\ref{dXim}), (\ref{dX0p}) and (\ref{dXip}) but which we recall here for convenience:
\begin{eqnarray}
&& V^0_1 + U^0_1 = v^0_-, \qquad V^i_1 - U^i_i = v^i_-, \nonumber \\
&& V^0_2 - U^0_2 = v^0_+,  \qquad V^i_2 + U^i_2 = v^i_+.
\label{vdef}
\end{eqnarray} 
We can set
\begin{equation}
\frac{v^{\mu}_+ + v^{\mu}_-}{2} = V^{\mu} \qquad \frac{v^{\mu}_+ -  v^{\mu}_-}{2} = U^{\mu}.
\label{VUdef}
\end{equation}
This amounts to combining together the two processes described above so that the asymmetry between space and time needed to obtain Eq.~(\ref{dxidxi}), which is the consequence of the request for covariance, disappears. Then, we combine the $V^{\mu}$ and $U^{\mu}$ 4-velocity fields to define the complex 4-velocity field
\begin{equation}
{\cal V}^{\mu} = V^{\mu} -i U^{\mu}.
\label{calV}
\end{equation}
The free Klein-Gordon equation follows from the reasoning described in Refs. \cite{CN04} and \cite{LN96} which is briefly recalled here.

From Eqs.~(\ref{VUdef}) and (\ref{calV}), we can define a complex derivative operator,
\begin{equation}
\frac{\dfr}{{\rm d}s} = {1\over 2} \left( \frac{{\rm d}}{{\rm d}s_+} + \frac{{\rm d}}{{\rm d}s_-} \right) - \frac{i}{2} \left(\frac{{\rm d}}{{\rm d}s_+} - \frac{{\rm d}}{{\rm d}s_-}\right),
\label{cdo}
\end{equation}
which, once applied to the position vector, gives the velocity field ${\cal V}^{\mu}$ of Eq.~(\ref{calV}).

Now, the total derivative of a fractal function $f(s, x^{\mu})$ of fractal dimension $D_f=2$ contains finite terms up to second order and can therefore be written
\begin{equation}
\frac{{\rm d}f}{{\rm d}s} = \frac{\partial f}{\partial s} + \frac{\partial f}{\partial x_{\mu}}  \frac{{\rm d}X_{\mu}}{{\rm d}s} + 
\frac{1}{2} \frac{\partial ^2 f}{\partial x_{\mu} \partial x_{\nu}} \frac{{\rm d}X_{\mu} {\rm d}X_{\nu}} {{\rm d}s}.
\label{totder}
\end{equation}
Substituting Eqs.~(\ref{dX0m}), (\ref{dXim}), (\ref{dX0p}) and (\ref{dXip}) into Eq.~(\ref{totder}) while using Eq.~(\ref{dxidxi}), we obtain two derivatives which we write as
\begin{equation}
\frac{{\rm d}f} {{\rm d}s_{\pm}} =  \left( \frac{\partial} {\partial s}  + v ^{\mu }_{\pm} \partial_{\mu }   \mp   {\cal D} \partial^{\mu } \partial_{\mu } \right) f.
\label{2diff}
\end{equation}
Since we consider only $s$-stationary functions, not explicitly depending on the proper time $s$, the complex derivative operator reduces to
\begin{equation}
\frac {\dfr}{{\rm d}s}   =  ({\cal V}^{\mu }  +  i  {\cal D} \partial^{\mu } ) \partial_{\mu }.
\label{comder}
\end{equation}
The plus sign in front of the Dalembertian comes from the choice of the metric signature.

To write the equation of motion, we use a generalized equivalence principle. We 
obtain thus a geodesic equation in terms of the complex derivative
\begin{equation}
\frac {\dfr {\cal V}_{\nu}}{{\rm d}s}=0.
\label{geoeq}
\end{equation}
We introduce a complex action according to
\begin{equation}
{\rm d}{\cal S}=\partial_{\nu}{\cal S}\; {\rm d}x^{\nu}=-mc \; {\cal V}_{\nu} \; {\rm d}x^{\nu}.
\label{comact}
\end{equation}
The complex four-momentum can thus be written as
\begin{equation}
{\cal P}_{\nu}=mc \; {\cal V}_{\nu}= -\partial_{\nu}{\cal S}.
\label{comfm}
\end{equation}
Since the complex action, ${\cal S}$, characterizes completely the dynamical state of the particle, we can introduce a complex wave function as
\begin{equation}
\psi    =   {\rm e}  ^{i{\cal S} /{\cal S}_0}.
\label{psi}
\end{equation}
It is linked to the complex four-velocity by Eq.~(\ref{comfm}), which gives
\begin{equation}
{\cal V} _{\nu } = \frac{i {\cal S}_0} {mc} \partial_{\nu }\ln\psi.
\label{comfv}
\end{equation}
Substituting Eqs.~(\ref{comder}) and (\ref{comfv}) into Eq.~(\ref{geoeq}), using the identity displayed as, e.g. (57) in Ref. \cite{CN04}, and with the choice ${\cal S}_0 = \hbar = 2mc{\cal D}$, we obtain
\begin{equation}
\partial^{\nu } \left( \frac{\partial^{\mu }\partial_{\mu }\psi}{\psi} \right) =  0,
\label{moteq}
\end{equation}
which can be integrated as the Klein-Gordon equation without electromagnetic field
\begin{equation}
\partial^{\mu } \partial_{\mu}\psi + \frac{m^2 c^2}{\hbar^2}\psi=0,
\label{KleinG}
\end{equation}
provided the integration constant is chosen equal to a squared mass term, $m^2c^2/\hbar^2$.

\subsection{The velocity field of the Dirac equation}
\label{Dir}

The results obtained above induce us to propose a new method for the construction, in the framework of SR, of the bi-quaternionic velocity field used to derive the bi-quaternionic Klein-Gordon equation, from which the Dirac equation is extracted \cite{CN03,CN04}. 

In addition to the definitions displayed in Eq.~(\ref{vdef}), we implement as follows new doublings of the velocity field. We add first, to the breaking of the symmetry ${\rm d} s \leftrightarrow - \,{\rm d} s$, the breaking of the local symmetry ${\rm d} x^{\mu} \leftrightarrow - \,{\rm d} x^{\mu}$. In the non-differentiable case, there is indeed no reason for ${\rm d} X^{\mu}$ to be equal to $-(- \,{\rm d} X^{\mu})$ as in the differentiable case. We can therefore add to the set of Eqs.~(\ref{dX01}), (\ref{dXi1}), (\ref{dX02}) and (\ref{dXi2}), a new set of equations
\begin{equation}
- {\rm d} X^0_1 = \bar{V}^0_1 {\rm d} s + \bar{U}^0_1 {\rm d} s + \bar{\eta}^0_1  \sqrt{2 {\cal D} {\rm d} s},
\label{mdX01}
\end{equation}
\begin{equation}
- {\rm d} X^i_1 = \bar{V}^i_1 {\rm d} s - \bar{U}^i_1 {\rm d} s + \bar{\eta}^i_1  \sqrt{- 2 {\cal D} {\rm d} s},
\label{mdXi1}
\end{equation}
\begin{equation}
- {\rm d} X^0_2 = \bar{V}^0_2 {\rm d} s - \bar{U}^0_2 {\rm d} s + \bar{\eta}^0_2  \sqrt{- 2 {\cal D} {\rm d} s},
\label{mdX02}
\end{equation}
\begin{equation}
- {\rm d} X^i_2 = \bar{V}^i_2 {\rm d} s + \bar{U}^i_2 {\rm d} s + \bar{\eta}^i_2  \sqrt{2 {\cal D} {\rm d} s}.
\label{mdXi2}
\end{equation}

We have now, with those of Eq.(\ref{vdef}) which we recall below, four velocity fields:
\begin{eqnarray}
&& V^0_1 + U^0_1 = v^0_-, \qquad V^i_1 - U^i_1 = v^i_-, \nonumber \\
&& V^0_2 - U^0_2 = v^0_+, \qquad V^i_2 + U^i_2 = v^i_+, \nonumber \\
&& \bar{V}^0_1 + \bar{U}^0_1 = \bar{v}^0_-, \qquad \bar{V}^i_1 - \bar{U}^i_1 = \bar{v}^i_-, \nonumber \\
&& \bar{V}^0_2 - \bar{U}^0_2 = \bar{v}^0_+, \qquad \bar{V}^i_2 + \bar{U}^i_2 = \bar{v}^i_+.
\label{vdef2}
\end{eqnarray}

We apply then the same reasoning to the $X^{\mu} \leftrightarrow - X^{\mu}$ symmetry, which is also broken in a fractal space-time. There is indeed no reason for $X^{\mu}$ to be equal to $- X^{\mu}$ and this induces a new doubling of the differential elements since ${\rm d} X^{\mu}$ has no reason to be equal to ${\rm d} (-X^{\mu})$ and $- {\rm d} X^{\mu}$ has no reason to be equal to $- {\rm d} (-X^{\mu})$. We can therefore add a new set of equations to Eqs.~(\ref{dX01}), (\ref{dXi1}), (\ref{dX02}), (\ref{dXi2}) and (\ref{mdX01})--(\ref{mdXi2}), which we write as
\begin{equation}
{\rm d} (-X^0_1) = \tilde{V}^0_1 {\rm d} s + \tilde{U}^0_1 {\rm d} s + \tilde{\eta}^0_1  \sqrt{2 {\cal D} {\rm d} s},
\label{mmdX01}
\end{equation}
\begin{equation}
{\rm d} (-X^i_1) = \tilde{V}^i_1 {\rm d} s - \tilde{U}^i_1 {\rm d} s + \tilde{\eta}^i_1  \sqrt{- 2 {\cal D} {\rm d} s},
\label{mmdXi1}
\end{equation}
\begin{equation}
{\rm d} (-X^0_2) = \tilde{V}^0_2 {\rm d} s - \tilde{U}^0_2 {\rm d} s + \tilde{\eta}^0_2  \sqrt{- 2 {\cal D} {\rm d} s},
\label{mmdX02}
\end{equation}
\begin{equation}
{\rm d} (-X^i_2) = \tilde{V}^i_2 {\rm d} s + \tilde{U}^i_2 {\rm d} s + \tilde{\eta}^i_2  \sqrt{2 {\cal D} {\rm d} s},
\label{mmdXi2}
\end{equation}
\begin{equation}
- {\rm d} (-X^0_1) = \breve{V}^0_1 {\rm d} s + \breve{U}^0_1 {\rm d} s + \breve{\eta}^0_1  \sqrt{2 {\cal D} {\rm d} s},
\label{bmdX01}
\end{equation}
\begin{equation}
- {\rm d} (-X^i_1) = \breve{V}^i_1 {\rm d} s - \breve{U}^i_1 {\rm d} s + \breve{\eta}^i_1  \sqrt{- 2 {\cal D} {\rm d} s},
\label{bmdXi1}
\end{equation}
\begin{equation}
- {\rm d} (-X^0_2) = \breve{V}^0_2 {\rm d} s - \breve{U}^0_2 {\rm d} s + \breve{\eta}^0_2  \sqrt{- 2 {\cal D} {\rm d} s},
\label{bmdX02}
\end{equation}
\begin{equation}
- {\rm d} (-X^i_2) = \breve{V}^i_2 {\rm d} s + \breve{U}^i_2 {\rm d} s + \breve{\eta}^i_2  \sqrt{2 {\cal D} {\rm d} s}.
\label{bmdXi2}
\end{equation}

We can therefore construct four new 4-velocity fields
\begin{eqnarray}
&& \tilde{V}^0_1 + \tilde{U}^0_1 = \tilde{v}^0_-, \qquad \tilde{V}^i_1 - \tilde{U}^i_1 = \tilde{v}^i_-, \nonumber \\
&& \tilde{V}^0_2 - \tilde{U}^0_2 = \tilde{v}^0_+, \qquad \tilde{V}^i_2 + \tilde{U}^i_2 = \tilde{v}^i_+, \nonumber \\
&& \breve{V}^0_1 + \breve{U}^0_1 = \breve{v}^0_-, \qquad \breve{V}^i_1 - \breve{U}^i_1 = \breve{v}^i_-, \nonumber \\
&& \breve{V}^0_2 - \breve{U}^0_2 = \breve{v}^0_+, \qquad \breve{V}^i_2 + \breve{U}^i_2 = \breve{v}^i_+.
\label{vdef3}
\end{eqnarray}

We define now a set of eight 4-velocity fields constructed as
\begin{eqnarray}
&& V^{\mu}= \frac{v^{\mu}_+ + v^{\mu}_-}{2}, \qquad U^{\mu} = \frac{v^{\mu}_+ - v^{\mu}_-}{2}, \nonumber \\
&& \bar{V}^{\mu} = \frac{\bar{v}^{\mu}_+ + \bar{v}^{\mu}_-}{2}, \qquad \bar{U}^{\mu} = \frac{\bar{v}^{\mu}_+ - \bar{v}^{\mu}_-}{2}, \nonumber \\
&& \tilde{V}^{\mu} = \frac{\tilde{v}^{\mu}_+ + \tilde{v}^{\mu}_-}{2}, \qquad \tilde{U}^{\mu} = \frac{\tilde{v}^{\mu}_+ - \tilde{v}^{\mu}_-}{2}, \nonumber \\
&& \breve{V}^{\mu} = \frac{\breve{v}^{\mu}_+ + \breve{v}^{\mu}_-}{2}, \qquad \breve{U}^{\mu} = \frac{\breve{v}^{\mu}_+ - \breve{v}^{\mu}_-}{2},
\label{VUpmdef}
\end{eqnarray}
which we combine to obtain four complex 4-velocity fields such as
\begin{eqnarray}
&& W^{\mu}_0 = V^{\mu} -i U^{\mu}, \nonumber \\
&& W^{\mu}_1 = \bar{V}^{\mu} -i \bar{U}^{\mu}, \nonumber \\
&& W^{\mu}_2 = \tilde{V}^{\mu} -i \tilde{U}^{\mu}, \nonumber \\
&& W^{\mu}_3 = \breve{V}^{\mu} -i \breve{U}^{\mu}.
\label{Wdef}
\end{eqnarray}

We can then construct the bi-quaternionic 4-velocity field such as
\begin{equation}
{\cal V}^{\mu} = W^{\mu}_0 + W^{\mu}_1 e_1 + W^{\mu}_2 e_2 + W^{\mu}_3 e_3,
\label{Vdir}
\end{equation}
where the $e_i$'s are the quaternion units. We see that this definition is more symmetric than the one provided in Refs. \cite{CN03} and \cite{CN04}, but its nicer property is to proceed directly from the scale relativistic first principles in a fully covariant and Lorentz invariant manner.

The derivative operator which, once applied to the position vector $X^{\mu}$, gives the velocity field defined by Eq.~(\ref{Vdir}) is
\begin{eqnarray}
\frac{\dfr}{{\rm d} s} &=& \frac{1}{2} \left(\frac{{\rm d}}{{\rm d} s}_+ + \frac{{\rm d}}{{\rm d} s}_-\right) -\frac{i}{2}\left(\frac{{\rm d}}{{\rm d} s}_+ - \frac{{\rm d}}{{\rm d} s}_-\right) \nonumber \\
&+& \left[ \frac{1}{2} \left(\frac{\bar{{\rm d}}}{{\rm d} s}_+ + \frac{\bar{{\rm d}}}{{\rm d} s}_-\right) -\frac{i}{2}\left(\frac{\bar{{\rm d}}}{{\rm d} s}_+ - \frac{\bar{{\rm d}}}{{\rm d} s}_-\right)\right] e_1 \nonumber \\
&+& \left[ \frac{1}{2} \left(\frac{\tilde{{\rm d}}}{{\rm d} s}_+ + \frac{\tilde{{\rm d}}}{{\rm d} s}_-\right) -\frac{i}{2}\left(\frac{\tilde{{\rm d}}}{{\rm d} s}_+ - \frac{\tilde{{\rm d}}}{{\rm d} s}_-\right)\right] e_2 \nonumber \\
&+& \left[ \frac{1}{2} \left(\frac{\breve{{\rm d}}}{{\rm d} s}_+ + \frac{\breve{{\rm d}}}{{\rm d} s}_-\right) -\frac{i}{2}\left(\frac{\breve{{\rm d}}}{{\rm d} s}_+ - \frac{\breve{{\rm d}}}{{\rm d} s}_-\right)\right] e_3.
\label{dop}
\end{eqnarray}

Substituting Eqs.~(\ref{dX01}), (\ref{dXi1}), (\ref{dX02}), (\ref{dXi2}), (\ref{mdX01})--(\ref{mdXi2}) and (\ref{mmdX01})--(\ref{bmdXi2}) into Eq.~(\ref{totder}) while using Eq.~(\ref{dxidxi}) written as $\langle \eta^{\mu}_{\pm} \eta^{\nu}_{\pm} \rangle = \mp \eta^{\mu \nu}$, we obtain a bi-quaternionic derivative operator which, owing to the fact that we consider only $s$-stationary functions, reduces to
\begin{equation}
\frac {\dfr}{{\rm d}s}   =  ({\cal V}^{\mu }  +  i  {\cal D} \partial^{\mu } ) \partial_{\mu }.
\label{bqder}
\end{equation}

We use now the generalized equivalence principle, as we did in Sec.~\ref{KG}. In this case, Eqs.~(\ref{geoeq})--(\ref{comfm}) still apply, while the bi-quaternionic wave function is expressed in terms of the action as
\begin{equation}
\psi^{-1} \partial_{\mu} \psi = \frac{i}{S_0} \partial_{\mu} {\cal S},
\label{bqpsi}
\end{equation}
using, in the left-hand side, the quaternionic product. This gives for the bi-quaternionic four-velocity, as derived from Eq.~(\ref{comfm}),
\begin{equation}
{\cal V}_{\mu}=i\frac{S_0}{mc} \psi^{-1} \partial_{\mu} \psi.
\label{bqv}
\end{equation}
Then, we proceed as in Sec.~\ref{KG}, using here the property of the quaternions,
\begin{equation}
\psi \, \partial _{\mu} \psi^{-1} = - (\partial _{\mu} \psi) \psi^{-1}, \quad
\psi^{-1} \partial _{\mu} \psi = - (\partial _{\mu} \psi^{-1}) \psi,
\label{comquat}
\end{equation}
and we obtain, after some calculations \cite{CN03,CN04},
\begin{equation}
\partial_{\mu}[(\partial^{\nu}\partial_{\nu} \psi) \psi^{-1}] = 0.
\label{bqdifeq}
\end{equation}
We integrate this four-gradient and take its right product by $\psi$ to obtain
\begin{equation}
\partial^{\nu}\partial_{\nu} \psi + C \psi = 0,
\label{bqKGeq}
\end{equation}
which we recognize as the Klein-Gordon equation for a free particle with mass $m$ (obtained for the choice of integration constant $m^2c^2/{\hbar}^2=C$), but now generalized to complex quaternions.

Then, we use a long-known property of the quaternionic formalism, which allows one
to obtain the Dirac equation for a free particle as a mere square root of 
the Klein-Gordon operator \cite{CN03,CN04,CL29,AC37}, i.e., by applying twice to the bi-quaternionic wave function $\psi$ the operator $\partial/c \partial t$ written as
\begin{equation}
\frac{1}{c} \frac{\partial}{\partial t} = e_3 \frac{\partial}{\partial x} e_2 + e_1 \frac{\partial}{\partial y} i + e_3 \frac{\partial}{\partial z} e_1 -i \frac{mc}{\hbar} e_3(\quad)e_3.
\label{Diracbiq}
\end{equation}
Using the properties of the Conway matrices \cite{AC37}, we write then Eq.~(\ref{Diracbiq}) as the non-covariant Dirac equation for a free fermion,
\begin{equation}
\frac{1}{c} \frac{\partial \psi}{\partial t} = - \alpha^k \frac{\partial \psi}{\partial x^k} - i \frac{mc}{\hbar} \beta \psi.
\label{Dirac}
\end{equation}

The isomorphism which can be established between the quaternionic and spinorial algebrae through the multiplication rules applying to the Pauli spin matrices allows us to identify the wave function $\psi$ with a Dirac bi-spinor.

It is also important to stress here that the breaking at the local level of the P symmetry, $X^i \leftrightarrow - X^i$, which transforms a dotted spinor into an undotted spinor and conversely \cite{LL72}, disappears at the global level thanks to the symmetry properties of the ${\cal V}^{\mu}$ velocity field due to its bi-quaternionic nature. The same reasoning applies to the T symmetry, $X^0 \leftrightarrow - X^0$. The breaking at the local level of this symmetry, which changes the wave function to its complex conjugate \cite{LL72}, disappears at the global level of the ${\cal V}^{\mu}$ velocity field thanks to its complex nature. As requested by relativistic quantum theory, these symmetries apply therefore to the full bi-spinor $\psi$ as defined by Eq.~(\ref{bqv}).


\section{Conclusions}
\label{concl}


We have generalized the new reasoning displayed in Paper I \cite{NC13}, leading to the natural emergence of the complex nature of the wave function in SR, to the full four-dimensional space-time. We have shown that the velocities become complex, then bi-quaternionic, because there are fundamental successive two-valuednesses of the mean derivatives due to non-differentiability.

The key point of the reasoning, which has been discussed in detail in Paper I \cite{NC13}, is the fact that ${\rm d} s$, which allows us to define the (+) and (-) derivatives of the fractal coordinates, can be viewed either as a differential element (in this case, it is algebraic) or as a resolution interval, $\varepsilon_s$. Then its sign looses its meaning and it appears in scale transformations in a logarithmic form. A particularly meaningful choice for the scale variable is $\varepsilon_s = |{\rm d} s|$. This special choice implies to work at the interface where the derivative takes its physical value whatever the scale. In a fractal space-time we are therefore led to consider a fractal 4-velocity field $V^{\mu}(x^{\mu}(s),s,{\rm d} s,\varepsilon_s) ={\rm d} X^{\mu}/{\rm d} s$, so that the differential element ${\rm d} X^{\mu}(x^{\mu}(s),s,{\rm d} s,\varepsilon_s)$ can be decomposed into a differentiable linear contribution and a non-linear fractal fluctuation.

We have thus presented here a detailed derivation of the way one can obtain a Lorentz invariant expression for the expectation value of the product of two independent fractal fluctuation fields in the sole framework of SR, while it had been up to now merely inferred from mathematical results obtained in stochastic mechanics. This relativistic and Lorentz invariant expression is a key step as regards the ability to obtain the Dalembertian operator appearing in the Klein-Gordon equation instead of the Laplacian operator showing in the Schr\"odinger equation.

For this purpose, we have considered that each fractal fluctuation behaves independently around its own mean equal to zero. We have thus justified to consider it as a stochastic process which we have chosen to be of Berstein type, since such a type includes a large class of stochastic processes among which Markov and anti-Markov processes. Following a reasoning inspired by Serva \cite{MS88}, we have considered two classes of stochastic processes that both contain a combination of Markovian and anti-Markovian processes: the first with a positive increment of the $s$ parameter for the time coordinate and negative increments of this $s$ parameter for the space coordinates, the second with a negative increment of the $s$ parameter for the time coordinate and positive increments of this $s$ parameter for the space coordinates. Combining these processes, we have been able to transform the Laplacian type expectation value of the product of two independent fractal fluctuation fields into a Dalembertian type one, implementing therefore in a robust manner the Lorentz invariance of SR.

We have thus been able to revisit the derivation of the two main equations of motion of relativistic quantum mechanics (the Klein-Gordon and Dirac equations) from successive symmetry breakings issuing naturally in the framework of the SR theory. \\


{\it Acknowledgments.} One of us, MNC, wishes to thank Daniel Sudarsky for an interesting discussion about Lorentz invariance.



\end{document}